# Scientific Utopia: II. Restructuring incentives and practices to promote truth over publishability

Brian A. Nosek

Jeffrey R. Spies

Matt Motyl

University of Virginia

Authors' note: Correspondence concerning this article may be sent to Brian Nosek, nosek@virginia.edu. We thank Yoav Bar-Anan, Roger Giner-Sorolla, Jesse Graham, Hal Pashler, Marco Perugini, N. Sriram, Victoria Stodden, and E.J. Wagenmakers for helpful comments. The authors have no financial interests concerning the content of this article.

Prepared for a special issue of *Perspectives on Psychological Science*

Version 1.2; May 25, 2012

Abstract

An academic scientist's professional success depends on publishing. Publishing norms emphasize novel, positive results. As such, disciplinary incentives encourage design, analysis, and reporting decisions that elicit positive results and ignore negative results. Prior reports demonstrate how these incentives inflate the rate of false effects in published science. When incentives favor novelty over replication, false results persist in the literature unchallenged, reducing efficiency in knowledge accumulation. Previous suggestions to address this problem are unlikely to be effective. For example, a journal of negative results publishes otherwise unpublishable reports. This enshrines the low status of the journal and its content. The persistence of false findings can be meliorated with strategies that make the fundamental but abstract accuracy motive – getting it right – competitive with the more tangible and concrete incentive – getting it published. We develop strategies for improving scientific practices and knowledge accumulation that account for ordinary human motivations and self-serving biases.

Abstract = 149 words

"The chief thing which separates a scientific method of inquiry from other methods of acquiring knowledge is that scientists seek to let reality speak for itself, and contradict their theories about it when those theories are incorrect... Scientific researchers propose hypotheses as explanations of phenomena, and design experimental studies to test these hypotheses via predictions which can be derived from them. These steps must be repeatable, to guard against mistake or confusion in any particular experimenter. … Scientific inquiry is generally intended to ... document, archive and share all data and methodology so they are available for careful scrutiny by other scientists, giving them the opportunity to verify results by attempting to reproduce them."

-- From http://en.wikipedia.org/wiki/Scientific_method, February 13, 2012

**A true story of what could have been**

Two of the present authors, Motyl and Nosek, share interests in political ideology. We were inspired by the fast growing literature on embodiment that demonstrates surprising links between body and mind (Markman & Brendl, 2005; Proffitt, 2006) to investigate embodiment of political extremism. Participants from the political left, right and center ($N$ = 1,979) completed a perceptual judgment task in which words were presented in different shades of gray. Participants had to click along a gradient representing grays from near black to near white to select a shade that matched the shade of the word. We calculated accuracy: How close to the actual shade did participants get? The results were stunning. Moderates perceived the shades of gray more accurately than extremists on the left and right ($p$ = .01). Our conclusion: political extremists perceive the world in black-and-white, figuratively *and* literally. Our design and follow-up analyses ruled out obvious alternative explanations such as time spent on task and a tendency to select extreme responses. Enthused about the result, we identified *Psychological Science* as our fall back journal after we toured the *Science, Nature,* and *PNAS* rejection mills. The ultimate publication, Motyl and Nosek (2012) served as one of Motyl's signature publications as he finished graduate school and entered the job market.

The story is all true, except for the last sentence; we did not publish the finding. Before writing and submitting, we paused. Two recent papers highlighted the possibility that research practices spuriously inflate the presence of positive results in the published literature (John, Loewenstein, & Prelec, 2012; Simmons, Nelson, & Simonsohn, 2011). Surely ours was not a case to worry about. We had hypothesized it, the effect was reliable. But, we had been discussing reproducibility, and we had declared to our lab mates the importance of replication for increasing certainty of research results. We also had an unusual laboratory situation. For studies that could be run through a web browser, data collection was very easy (Nosek et al., 2007). We could not justify skipping replication on the grounds of feasibility or resource constraints. Finally, the procedure had been created by someone else for another purpose, and we had not laid out our analysis strategy in advance. We could have made analysis decisions that increased the likelihood of obtaining results aligned with our hypothesis. These reasons made it difficult to avoid doing a replication. We conducted a direct replication while we prepared the manuscript. We ran 1,300 participants, giving us .995 power to detect an effect of the original effect size at alpha = .05. The effect vanished ($p$ = .59).

Our immediate reaction was “why the #&@! did we do a direct replication?” Our failure to replicate is not definitive that the original effect is false, but it raises enough doubt to make reviewers recommend against publishing. Any temptation to ignore the replication and publish the original only was squashed by the fact that our lab mates knew we ran a replication. We were accountable to them. The outcome – a dead or delayed paper – is unfortunate for our career advancement, particularly Motyl’s as he prepared for the job market.

Incentives for surprising, innovative results are strong in science. Science thrives by challenging prevailing assumptions and generating novel ideas and evidence that push the field in new directions. We cannot expect to eliminate the disappointment that we felt by “losing” an exciting result. That is not the problem, or at least not one for which the fix would improve scientific progress. The real problem is that the incentives for publishable results can be at odds with the incentives for accurate results. This produces a conflict of interest. The conflict may increase the likelihood of design, analysis, and reporting decisions that inflate the proportion of false results in the published literature.[1] The solution requires making incentives for *getting it right* competitive with the incentives for *getting it published*. Without that, the lesson that we could take away from our experience with "Political extremists do not perceive shades of gray, literally" is to never, ever do a direct replication again. The purpose of this article is to make sure that such a lesson does not stick.

**How evaluation criteria can increase the false result rate in published science**

Publishing is "the very heart of modern academic science--at levels ranging from the epistemic certification of scientific thought to the more personal labyrinths of job security, quality of life, and self-esteem" (Mahoney, 1985, pp. 30). Publication influences hiring, salary, promotion, tenure, and grant decisions (Miller & Serzan, 1984; Peters & Ceci, 1982). It is also a criterion for the evaluation and ranking of departments and universities (Ball, 2005; Haslam et al., 2008; Nosek et al., 2010; Ostriker, Holland, Kuh, & Voytuk, 2009; Phillips, 2007). Expectations for publishing have expanded to faculty at institutions that do not have graduate research programs, to graduate students seeking employment, and even to undergraduates applying to top programs for graduate study. With an intensely competitive job market, the demands for publication might seem to suggest a specific objective for the early-career scientist: publish as many articles as possible in the most prestigious journals that will accept them (Martin, 1992; Sovacool, 2008).[2]

[1] We endorse a perspectivist approach to science (McGuire, 2004) – the idea that all claims may be true given the appropriate conditions. In this article, when we say *true* we mean the truth of the claim as it is stated, usually conceived as the ordinal relationship between conditions, effects, or direction of correlation (Frick, 1996). The general truth value of a claim is established by expressing the limiting conditions under which it is true. Without expressing those conditions, the claim is likely to be false or, at best, partly true.

[2] Later, we will argue that this is more the perceived than the real formula for success. For now, we are dealing with perception, not reality.

*Some things are more publishable than others*. Even if a researcher conducts studies competently, analyzes the data effectively, and writes up the results beautifully, there is no guarantee that the report will be published. Part of the process - peer review - is outside of the researcher's control. In the social and behavioral sciences, rejection rates of 70-90% by journals are common (American Psychological Association, 2011; Zuckerman & Merton, 1971). High demand for limited space means that authors must strive to meet all publishing criteria so that an editor will do the unusual act of accepting the manuscript. As such, success in publishing is partly a function of social savvy of knowing what is publishable, and empirical savvy in obtaining publishable results.

**A disconnect between what is good for scientists and what is good for science**

On its own, the fact that publishing is essential to success is just a fact of the trade. Running faster defines better sprinters; conducting more high-impact research defines better scientists. The research must be published to have impact. And yet, publishing is also the basis of a conflict of interest between personal interests and the objective of knowledge accumulation. The reason? *Published* and *true* are not synonyms. To the extent that publishing itself is rewarded, then it is in scientists' personal interests to publish, regardless of whether the published findings are true (Hackett, 2005; Martin, 1992; Sovacool, 2008).

The present authors have accuracy motives - to learn and publish true things about human nature. We also have professional motives - to succeed and thrive professionally. Our incentives for professional success can be at odds with scientific practices that improve confidence in the truth of findings. Strong professional motives produce motivated reasoning to arrive at the conclusion we desire, even at the expense of accuracy (Kunda, 1990).

At the extreme, we could lie: make up findings or deliberately alter results. However, detection of such behavior destroys the scientist's reputation. This is a strong incentive against it, and - regardless of incentives - most resist such behavior because it is easy to identify as wrong (Fanelli, 2009).[3] We have enough faith in our values to believe that we would rather fail than fake our way to success. Less simple to put aside are ordinary practices that can increase the likelihood of publishing false results, particularly those practices that are common, accepted, and even appropriate in some circumstances. Because we have directional goals for success, we are likely to bring to bear motivated reasoning to justify research decisions in the name of accuracy, when they are actually in service of career advancement (Fanelli, 2010a). Motivated reasoning is particularly influential when the situation is complex, the available information is ambiguous, and legitimate reasons can be generated for multiple courses of action (Bersoff, 1999; Boiney, Kennedy, & Nye, 1997; Kunda, 1990).

Motivated reasoning can occur without intention. We are more likely to be convinced that our hypothesis is true, accepting uncritically when it is confirmed and scrutinizing heavily when it is not (Bastardi, Uhlmann, & Ross, 2011; Ditto & Lopez, 1992; Lord, Ross, & Lepper, 1979; Pyszczynski & Greenberg, 1987; Trope & Bassok, 1982). With flexible analysis options, we are more likely to find the one that is more publishable to be more reasonable and defensible than others that are less publishable (Simmons et al., 2011; Wagenmakers, Wetzels, Borsboom, & van der Maas, 2011). Once we obtain an unexpected result, we are likely to reconstruct our histories and perceive the outcome as something that we could have, even did, anticipate all along – converting a discovery into a confirmatory result (Fischoff, 1977; Fischoff & Beyth, 1975). And, even if we resist those reasoning biases in the moment, after a few months, we might simply forget the details, whether we: had hypothesized the moderator, had good justification for one set of exclusion criteria compared to another, and had really thought that the one dependent variable that showed a significant effect was the key outcome. Instead, we might remember the gist of what the study was and what we found (Reyna & Brainerd, 1995). Forgetting the details provides an opportunity for reimagining the study purpose and results to recall and understand them in their best (i.e., most publishable) light. The reader may, as we do, recall personal examples of such motivated decisions – they are entirely ordinary products of human cognition.

[3] Notably, it is difficult to detect deliberate malfeasance. The three most prominent cases in psychology's recent history - Karen Ruggiero, Marc Hauser, and Diederik Stapel - were not identified by disconfirmation of their results in the published literature (though, in Hauser's case, there was some public skepticism for at least one result). The misbehavior was only identified because colleagues - particularly junior colleagues - took considerable personal risk by voicing concerns about the internal practices of the laboratory.

**Novelty and positive results are vital for publishability, but not for truth**

The primary objective of science is to accumulate knowledge about nature. Learning something new advances that goal; reaffirming something known does not. As Schmidt (2009) noted, "within the social sciences, only the discovery of a new fact is credited" (Schmidt, 2009, p. 95; see also Lindsay & Ehrenberg, 1993). Innovation in ideas, methods and evidence are the basis for scientific advancement. As such, successful scientists are those that can identify a productive program of research that reveals facts about nature.

Related to this, direct replication of another's study procedures to confirm the results is uncommon in the social sciences (Collins, 1985; Mahoney, 1985; Schmidt, 2009). Neuliep and Crandall (1990; see also Madden, Easley, & Dunn, 1995; Neuliep and Crandall, 1993) found that 94% of journal editors agreed that "replication studies were not included as examples of research encouraged for submission in the editorial policy. . ." (p. 87) and a large majority preferred to publish new findings rather than replications because the latter were deemed "not newsworthy" and a "waste of space."

These comments are revealing about the relative valuation of new ideas versus truth. Publishing a result does not make it true. Many published results have uncertain truth value. Dismissing a direct replication as "we already knew that" is misleading; the actual criticism is "someone has already claimed that." The former indicates that the truth value is known, the latter indicates that someone has had the idea and perhaps provided some evidence. Replication is a means of increasing the confidence in the truth value of a claim. Its dismissal as a waste of space incentivizes novelty over truth. As a consequence, when a false result gets into the published literature, it is difficult to expel. There is little reinforcement for conducting replications to affirm or reject the validity of prior evidence and few consequences for getting it wrong. The principal incentive is publication.

Further, in the dominant model of null hypothesis significance testing (Cohen, 1994; Nickerson, 2000; Rosnow & Rosenthal, 2009; Rozeboom, 1960), the nominal false-positive rate of alpha = .05 has become a *de facto* criterion for publishing. Like publishing a result, achieving a positive result does not mean that the effect is true, nor does it indicate the probability of its truth (Bakan, 1966; Cohen, 1994; Frick, 1996; Oakes, 1986). Moreover, most published results across scientific disciplines, and over 90% for psychology in particular, are positive effects (Fanelli, 2010b, 2012). This appears to have been true for more than 50 years (Sterling, 1955; Sterling, Rosenbaum & Weinkam, 1995). If anything, the rate of positive results is increasing (Fanelli, 2012). Greenwald (1975) showed that psychologists perceive a bias against negative (null) results and are less likely to continue pursuing or report negative results, and that journals are less likely to publish negative as compared to positive results. As a consequence, negative results are less likely to appear in the literature.

In summary, the demands for novelty and positive results create incentives for (a) generating new ideas rather than pursuing additional evidence for or against ideas suggested previously, (b) reporting positive results and ignoring negative results (Fanelli, 2012; Greenwald, 1975; Ioannidis & Trikalinos, 2007; Rosenthal, 1979), and (c) pursuing design, reporting, and analysis strategies that increase the likelihood of obtaining a positive result in order to achieve publishability (Fanelli, 2010a; Ioannidis, 2005; John et al., 2012; Simmons et al., 2011; Wicherts, Bakker, & Molenaar, 2011; Wong, 1981; Young, Ioannidis, & Al-Ubaydli, 2008). This paints a bleak picture of the incentive structures in science. Simultaneously, we believe that a lot of very good science gets done and does so with incentives and practices that facilitate knowledge accumulation. We believe that "good behaviors" can be promoted further with some adaptations to normative scientific culture and practices and incentives structures that promote and sustain those practices. Before discussing solutions, we briefly review some of the practices that may interfere with efficiency in knowledge accumulation.

**Practices that can increase the proportion of false results in the published literature**

Other contributions have detailed a variety of practices that can increase publishability but might simultaneously decrease validity (Fanelli, 2010a; Giner-Sorolla, 2012; Greenwald, 1975; Ioannidis, 2005; John et al., 2012; Kerr, 1998; Martinson, Anderson, & Devries, 2005; Rosenthal, 1979; Simmons et al., 2011; Sovacool, 2008; Young et al., 2008). The following are practices that are justifiable sometimes, but can also increase the proportion of published false results: [1] leverage chance by running many low-powered studies, rather than a few high-powered ones[4] (Ioannidis, 2005); [2] uncritically dismiss "failed" studies as pilot tests or due to methodological flaws, but uncritically accept "successful" studies as methodologically sound (Bastardi et al., 2011; Lord, Ross, & Lepper, 1979); [3] selectively report studies with positive results and not studies with negative results (Greenwald, 1975; John et al., 2012; Rosenthal, 1979), or selectively report "clean" results (Begley & Ellis, 2012; Giner-Sorolla, 2012); [4] stop data collection as soon as a reliable effect is obtained (John et al.,2012; Simmons et al., 2011); [5] continue data collection until a reliable effect is obtained (John et al.,2012; Simmons et al., 2011); [6] include multiple independent or dependent variables, report the subset that "worked" (Ioannidis, 2005; John et al., 2012; Simmons et al., 2011); [7] maintain flexibility in design and analytic models including the attempt of a variety of data exclusion or transformation methods, report a subset (Gardner, Lidz, & Hartwig, 2005; Ioannidis, 2005; Martinson et al., 2005; Simmons et al., 2011); [8] report a discovery as if it had been the result of a confirmatory test (Bem, 2003; John et al., 2012; Kerr, 1998); and, [9] once a reliable effect is obtained, do not do a direct replication (Collins, 1985; Schmidt, 2009; see also Motyl & Nosek, 2012 [alternate timeline]).

[4] Reasonable justification: I am doing innovative research on a new phenomenon. Our resources for data collection are limited. It would be a poor use of resources to invest heavily if there is no effect to detect, or if I am pursuing it the wrong way. Unreasonable consequence: If the effect being investigated does not exist, the best way to obtain a significant result by chance is to run multiple small sample studies. If the effect being investigated does exist, the best way to confirm it is to run a single high-powered test.

The disinterest in replication is striking given its centrality to science. The scientific method differentiates itself from other approaches by publicly disclosing the basis of evidence for a claim. In Roger Bacon's cycle of scientific research - observation, hypothesis, experimentation, and verification - *disclosure* is critical for the fourth step (Bacon, 1267/1859). Other scientists must be able to independently replicate and verify, qualify or disconfirm the original scientist's results. This allows scientists to work independently toward a shared objective – accumulating knowledge – without relying on accuracy or trust in any single source. In principle, open sharing of methodology means that the entire body of scientific knowledge can be reproduced by anyone. This democratizing function for acquiring knowledge made *replication* a central principle of the scientific method from before Bacon to the present (e.g., al Haytham, 1021 as translated by Sabra, 1989; Jasny, Chin, Chong, & Vignieri, 2011; Kuhn, 1962; Lakatos, 1978; Popper, 1934; Rosenthal, 1991; Schmidt, 2009).[5] Replication is so central to science that it may serve as a "demarcation criterion between science and nonscience" (Braude, 1979, p. 2). Even so, direct replications are seldom pursued in the behavioral sciences perhaps because they are deemed unpublishable as failures of novelty (Collins, 1985; Reid, Soley, & Wimmer, 1981; Schmidt, 2009).

Many of the behaviors that can increase the rate of false results are common because there are good reasons to do them sometimes. The key challenge, however, is that there are good reasons to do them *sometimes*. For example, when wading into a new phenomenon, having multiple dependent variables can be a more efficient use of resources because there is little existing knowledge for informing which outcome might be affected. Simultaneously, this practice inflates the likelihood of false positives. This fact merely increases the importance of replicating the initial finding and disclosing that the initial study included multiple dependent variables, and why. The multiple dependent variable design decision is sensible, not embarrassing. The disclosure just provides evaluators with an accurate basis for computing their confidence in the results (Simmons et al., 2011).

[5] An exception is the scientific anarchist Feyerabend (1975) who rejected the notion that there were any universal methodological rules for the scientific method, and that science had no special status for identifying "objective" truths than any other approach.

The consequences of the publishability-improving practices listed above can be severe. Ioannidis (2005) gave his review of reproducibility a dire title: "Why most published research results are false." Greenwald (1975) estimated the false positive (Type I error) rate at 30% based only on whether researchers submit and editors accept negative findings. Bayer HealthCare reported that of 67 attempts to reproduce published findings in fields of oncology, women's health, and cardiovascular disease, about 25% of the replications clearly reproduced the published evidence (Prinz, Schlange, & Asadullah, 2011). This low rate was not attributable to publishing journal prestige, closeness of the replication attempt, or the subdiscipline of investigation. Likewise, Begley and Ellis (2012) reported an effort by Amgen to replicate 53 landmark studies of basic research for cancer treatment. Just 6 (11%) of the replications confirmed the original, published result. They noted, "Some non-reproducible preclinical papers had spawned an entire field, with hundreds of secondary publications that expanded on elements of the original observation, but did not actually seek to confirm or falsify its fundamental basis" (p. 532). Finally, an informal assumption among venture capital firms for biomedical research is that more than 50% of published studies from academic laboratories cannot be replicated in industrial laboratories (Osherovich, 2011). In these latter cases, the industrial laboratories pursued replications of academic research because there are considerable incentives for doing so. Investing hundreds of thousands of dollars on a new treatment that is ineffective is a waste of resources and an enormous burden to patients in experimental trials. By contrast, for academic researchers there are few consequences for being wrong. If replications get done and the original result is irreproducible nothing happens.

**Strategies that are not sufficient to stop the proliferation of false results**

False effects interfere with knowledge accumulation. If common scientific practices are increasing the rate of false effects, then changing some practices could improve efficiency in scientific progress. Of course, science is progressing and knowledge is accumulating. Revisions to scientific practices ought not to disrupt those practices that are working well. They should selectively target dysfunctional incentives and practices. Before turning toward our suggested improvements, we briefly review innovations that have been suggested or attempted that are not, in our view, sufficient to address the proliferation of false effects.

*Conceptual replication*. While direct replication is rare in the social and behavioral sciences, conceptual replication is more common (Collins, 1985; Schmidt, 2009). As opposed to direct replication, which reproduces the conditions of the original demonstration as faithfully as possible, conceptual replication involves deliberately changing the operationalization of the key elements of the design such as the independent variable, dependent variable, or both. Conceptual replications allowing abstraction of the explanation for an effect from the particulars of a given operationalization to the theoretical variable that operationalizations attempt to manipulate or assess (Schmidt, 2009). This is vitally important in science when the constructs of interest are unobservable (Edge, 1985). Demonstrating the same effect with multiple operationalizations provides confidence in its conceptual interpretation.

While often essential for theoretical understanding, conceptual replication is not an effective replacement for direct replication. As Schmidt (2009, p. 95) put it, “Whereas a direct replication is able to produce facts, a conceptual replication may produce understanding.” Because features of the original design are changed deliberately, conceptual replication is only used to confirm (and abstract) the original result, not to disconfirm it. A successful conceptual replication is used as evidence for the original result; a failed conceptual replication is dismissed as not testing the original phenomenon (Braude, 1979).[6] As such, using conceptual replication as a replacement for direct replication is the scientific embodiment of confirmation bias (Nickerson, 1998).

*The mythology of science as self-correcting*. Science is self-correcting (Merton, 1942, 1973). If a claim is wrong, eventually new evidence will accumulate to show that it is wrong and scientific understanding of the phenomenon will change. This is part of the promise of science – following the evidence where it leads, even if it is counter to present beliefs (see opening quotation of this article). We do believe that self-correction occurs. Our problem is with the word "eventually." The *myth of self-correction* is recognition that once published there is no systemic ethic of confirming or disconfirming the validity of an effect. False effects can remain for decades, slowly fading or continuing to inspire and influence new research (Prinz et al., 2011). Further, even when it becomes known that an effect is false, retraction of the original result is very rare (Budd, Sievert, Schultz, 1998; Redman, Yarandi & Merz, 2008). Researchers that do not discover the corrective knowledge may continue to be influenced by the original, false result. We can agree that the truth will win *eventually*, but we are not content to wait.

*Journals devoted to publishing replications or negative results*. An obvious strategy for addressing the lack of interest in publishing negative results or replications is to start journals devoted to publishing them (e.g., http://www.jasnh.com/; http://www.jnr-eeb.org/; http://www.journalofnullresults.com/). Unfortunately, we believe this model is doomed to fail. Defining a journal based on negative results or replications is self-defining it as a low importance outlet. For example, the *Journal of Personality and Social Psychology* has an explicit policy against publishing direct replications, communicating their lower status (Aldhous, 2011). It is not in authors' interest to publish in a journal that is defined as publishing articles that no other journal will publish.

[6] In reality, conceptual and direct replications exist on a continuum rather than being discrete entities (Schmidt, 2009). There is no such thing as an "exact" replication outside of simulation research because the exact conditions of the original investigation can never be duplicated. Direct replication therefore means that the original conditions are reproduced such that there is no reason to expect a different result based on its present interpretation. If sample, setting, or procedural factors are essential, then those must be specified in order to have a proper theoretical understanding. As such, among other reasons, a failure to replicate could mean that the conditions necessary to elicit the original result are not yet understood (see Open Science Collaboration [2012a] for more about possible interpretations of a failure to replicate). Further, deciding that a conceptual replication (whether successful or unsuccessful) tests the same phenomenon as an original result is usually a qualitative assessment rather than an empirical one.

*Education campaigns emphasizing the importance of replication and reporting negative results*. If editors, reviewers, and authors are not interested in publishing negative results or replications, then perhaps they could be educated that negative results and replications are important and should be valued like other contributions. This will not work either. Or, more accurately, it has not worked already. These issues have been prominent topics of discussion of methodology for more than three decades with little change in daily practices (Greenwald, 1975; Rosenthal, 1979). There is, for example, little disagreement that the file drawer effect is a bad thing.

Publishing practices are hard to change because innovative research is more important than replication research and negative results. Innovation is the key driver of scientific progress. Publishing has significant resource constraints. Most journals have page constraints for the number of articles they can publish each year, and they receive many more submissions than can be published even if they wished to do so (Nosek & Bar-Anan, 2012). When faced with the choice between accepting an article reporting a new innovation versus an article reporting a replication or negative result, editors and reviewers will usually select the former. Further, editors have the luxury of demanding aesthetically "clean" results rather than tolerating the reality of many research enterprises as untidy affairs (Giner-Sorolla, 2012; Kaiser, 2012). And, because demand for the limited space is so high, there is no shortage of articles reporting innovative ideas in neat packages to choose from.

*Increasing expectations of reviewers to catch motivated reasoning and other signs of false results*. Reviewers and editors are gatekeepers for scientific publishing. If they are not convinced that the manuscript meets the standards for the journal, the paper is rejected. It is conceivable then, to ask reviewers to scrutinize research more carefully for signs of false results (Schroter et al., 2008). Of the suggestions in this section, we believe that this one is the most reasonable for two reasons. First, reviewers are already very discriminating. It is likely that editors and reviewers are effective to some extent already at recognizing signals of false results. And, second, some evaluation suggestions would be easy to implement, such as the checklist suggested by Simmons and colleagues (2011).

Given the existing system, the peer review process offers the best method presently for identifying potentially false results, other than the diligence of the authors themselves. Nonetheless, we perceive this as a partial solution for three reasons. First, peer reviewers are volunteers. They already work hard for little to no reward. Requiring more than the simple suggestions is asking a lot of people that have already done more than their due. Second, reviewing is hard work. Even diligent reviewers miss lots of errors (Schroter et al., 2008; Wicherts et al., 2011). And, third, peer reviewers only review the summary report of the research, not the research itself. The report is the authors' perspective on how the research happened and what it means. Most of the actual research process never makes it into the pages of the report; measures, methods, and analysis strategies are summarized. Further, standard practice is to present a strong narrative of what the reader should learn from the research rather than describing how the research and learning occurred (Bem, 2003). The latter would likely be more revealing of potential problems than the former.

*Raising the barrier for publication.* A related alternative for gatekeepers is to raise the standards for publication by requiring reports to include replications of new findings (Begley & Ellis, 2012; Hewitt, 2012). In one sense, this solution would be effective. If editors did not publish articles without replications, then published articles would have replications. There are some cases, such as the opening anecdote, in which requiring a replication is reasonable – the effect is provocative and data collection is straightforward. However, we believe that requiring replication as a blanket editorial policy could be an impediment for scientific progress. For one, the standards for publication are already extremely high, especially in the social and behavioral sciences. Manuscripts take years to get published and they are often reviewed at multiple journals by multiple review teams (Nosek & Bar-Anan, 2012). Increasing expectations would further extend this process and put more demand on editors and reviewers – particularly because the pressure to publish will remain so strong for authors.

Moreover, requiring replication of everything could stifle risk-taking and innovation. In some domains, requiring replication is not an onerous requirement because data collection is easy. But, in other domains, it is difficult or even impossible to conduct a direct replication because of the resource demands or unique opportunities for data collection (Elms, 1975; Lykken, 1968). If replication were essential for every new phenomenon, then researchers might be disinclined to pursue new and challenging ideas to ensure publishability of what they produce. Innovation requires taking risks. That means that innovators can be wrong, perhaps frequently. The problem is not that false results get into the literature. The problem is that they stay in the literature. The best solutions would encourage innovation and risk-taking, but simultaneously reward confirmation of existing claims.

**Strategies that will accelerate the accumulation of knowledge**

In our view, the key for improving the efficiency of knowledge accumulation is to capitalize on existing motivations to be accurate and to reduce the emphasis on publication itself as the mechanism of achievement and advancement. Scientists have strong accuracy motivations. And, in the long run, *getting it right* has a higher payoff than *getting it published*. However, the goal to publish is immediate, palpable, and concrete; the goal to be accurate is distal and abstract. As a consequence, the short-term payoffs of publishing can be inordinately influential (Liberman & Trope, 1998; Trope & Liberman, 2003), particularly for early-career scientists for whom there is relative urgency for markers of achievement. To address this, the conditions of daily practice must elevate the importance of the more abstract, longer-term goals in comparison to the persisting importance of the concrete, shorter-term goals. In this section, we suggest new or altered practices to meet these objectives.

*Promoting and rewarding paradigm-driven research*. While conceptual replication is only used to confirm prior results, another relatively common research strategy – paradigm-driven research – can be used for both confirming and disconfirming prior results. Paradigm-driven research accumulates knowledge by systematically altering a procedure to investigate a question or theory, rather than varying many features of the methodology – by design or by accident. This offers an opportunity to incorporate replication and extension into a single experimental design (Roediger, 2012). Paradigm-driven research balances novelty and replication by building new knowledge using existing procedures. Effective use of this approach requires development of standards, sharing and reuse of materials, and deliberate alteration of design rather than wholesale reinvention. For example, the Deese-Roediger-McDermott paradigm for studying false memories (Roediger & McDermott, 1995) has been adapted to examine how aging (Butler, McDaniel, Dornburg, Price & Roediger, 2004), mood (Storbeck & Clore, 2005) and expectations (Schacter, Israel, & Racine, 1999) influence the frequency of false memories (see Gallo, 2010 for a review). The subsequent findings reinforce the original results through direct replication and extend those findings by identifying moderating influences, mechanisms, and boundary conditions. A paradigm-driven approach provides confidence in the validity of an effect (or doubt if it fails to replicate), while simultaneously extending knowledge in new directions.

It is easy to do more paradigm-driven research if authors make their paradigms available to others. The primary risk of paradigm-driven research is that research questions can evolve to being about the method itself rather than the theory that the method is intended to address. Using a single methodology for a theoretical question can reify idiosyncratic features of that methodology as being the phenomenon. This is where conceptual replication provides substantial added value. Paradigm-driven research provides confidence in the accuracy of findings. Conceptual replication ensures that the findings are theoretically general, not methodologically idiosyncratic.

*Author, reviewer, and editor checklists.* In the prior section we expressed some doubt in raising expectations of reviewers for catching errors with one exception – easy to implement checklists such as that suggested by Simmons and colleagues (Simmons et al., 2011). Checklists are an effective means of improving the likelihood that particular behaviors are performed and performed accurately (Gawande, 2009). Authors already follow some checklist-like requirements such as APA or MLA formatting. It is easy to conceive of statistical and disclosure checklists for authors and editorial teams. For example, CONSORT has a 25-item checklist describing minimum standards for reporting Randomized Controlled Trials (http://www.consort-statement.org/). Checklists can ensure disclosure of obvious items that are sometimes forgotten: sample sizes, statistical tests, effect sizes, and covariates included in analysis. They can also define best practices and methodological standards for domain specific applications.

Why are checklists needed? The most straightforward reason is that key information is left out with stunning frequency, and advisable methodological practices are not identified "naturally" or systematically in the review process. For example, the value of reporting effect sizes has been widely disseminated (Cohen, 1962, 1969, 1992; Wilkonson and Task Force on Statistical Inference, 1999). Nonetheless, reporting effect sizes has only become common in recent history, and is still not standard practice. A checklist requiring their inclusion before publication would change this. Further, Bouwmeester and colleagues examined 71 prediction studies from high-impact medical journals and found pervasive methodological shortcomings in design, reporting, and analysis decisions such as clear specification of predictor and outcome variables, description of participant exclusion criteria, and handling of missing values (Bouwmeester et al., 2012). They concluded that "The majority of prediction studies in high impact journals do not follow current methodological recommendations, limiting their reliability and applicability." High standards for publication do not translate into *specific* standards for reporting. Authors, reviewers and editors examination of each article is almost entirely *ad hoc*. Societies, journals, and individuals could maintain simple checklists of standard requirements to prevent errors and improve disclosure.

*Challenging mindsets that sustain the dysfunctional incentives*. Earlier we stated: "With an intensely competitive job market, the demands for publication might seem to suggest a specific objective for the early-career scientist: publish as many articles as possible in the most prestigious journals that will accept them." While this is a common perception, particularly among early-career scientists, we also believe that there are good reasons – though not yet sufficient evidence – to challenge it. For example, the first author regularly presents to graduate students summary data of the short list from a past search for an assistant professor in Psychology at the University of Virginia. For this particular search, more than 100 applications were received. The Table presents the 11 applicants that made it to the short-list. All short-list candidates had at least 4 publications and at least 1 first-authored publication. Based on publication numbers, there are clear stand-outs from this group such as the postdoc with 35 publications, an assistant professor with 21 publications, and a graduate student with 10 publications. Further, these candidates published in prestigious outlets. However, none of these three were selected as a finalist. In fact, two of the three interviewed candidates were among the least productive on the short-list.

This anecdote suggests that some degree of publishing productivity is essential to get into the pool of competitive candidates but, after that, other factors are more important for getting the job. Without sufficient evidence, we speculate that publication numbers and journal prestige heuristics do play a role in initial selection from a large hiring pool, and then play a much smaller role when the pool is narrowed and the hiring committees can look deeply at each candidate. At that point, the committees can invest time to examine quality, potential impact, and direction of the research agenda. In tenure and promotion cases, the depth of processing ought to be even more acute as it is a detailed review of a single candidate's record.

This conclusion is based on anecdotal data. Early-career scientists would get useful information from a systematic review of the degree to which publication numbers and journal prestige predict hiring and promotion. Multiple departments could pool and share evidence. The aggregate data might confirm the prevailing perception that publication numbers and journal prestige are the key drivers for professional success or, as we believe, they would illustrate notably weaker predictive validity when the evaluation committee has resources to examine each record in detail.

*Metrics to identify what is worth replicating*. Even if valuation of replication increased, it is not feasible – or advisable – to replicate everything. The resources required would undermine innovation. A solution to this is to develop metrics for identifying *Replication Value* (RV)– what effects are more worthwhile to replicate than others? The Open Science Collaboration (2012b) is developing an RV metric based on the citation impact of a finding and the precision of the existing evidence of the effect. It is more important to replicate findings with a high RV because they are becoming highly influential and yet their truth value is still not precisely determined. Other metrics might be developed as well. Such metrics could provide guidance to researchers for research priorities, to reviewers for gauging the "importance" of the replication attempt, and to editors who could, for example, establish an RV threshold that their journal would consider as sufficiently important to publish in its pages.

*Crowdsourcing replication efforts*. Individual scientists and laboratories may be interested in conducting replications, but not have sufficient resources available for them. It may be easier to conduct replications by crowdsourcing them with multiple contributors. For example, in 2011, the Open Science Collaboration began investigating the reproducibility of psychological science by identifying a target sample of studies from published articles from 2008 in three prominent journals – *Journal of Personality and Social Psychology, Journal of Experimental Psychology: Learning, Memory, and Cognition*, and *Psychological Science* (Carpenter, 2012; Yong, 2012). Individuals and teams selected a study from the eligible sample and followed a standardized protocol. In the aggregate, the results were intended to facilitate understanding of the reproducibility rate and factors that predict reproducibility. Further, as an open project, many collaborators could join and make small contributions that accumulate into a large-scale investigation. The same concept can be incorporated into replications of singular findings. Some important findings are difficult to replicate because of resource constraints. Feasibility could be enhanced by spreading the data collection effort across multiple laboratories.

*Journals with peer review standards focused on the soundness, not importance, of research.* The basis of rejection for much research is that it does not meet the criterion of being sufficiently "important" for the journal considering it. Many manuscripts are rejected on this criterion, even if the reviewers identify the research as sound and reported effectively. Despite evidence of the unreliability of the review process for evaluation and identifying importance (Bornmann, Mutz, & Daniel, 2010; Cicchetti, 1990; Gottfredson, 1978; Marsh & Ball, 1989; Marsh, Jayasinghe, & Bond, 2008; Petty, Fleming & Fabrigar, 1999; Peters & Ceci, 1982; Whitehurst, 1984), this is a reasonable criterion given that journals have limited space and desires to be prestigious outlets. However, in the digital age, page limits are an anachronism (Nosek & Bar-Anan, 2012). Digital journal *PLoS ONE* (http://plosone.org/) publishes research from any area of scientific inquiry. Peer review at *PLoS ONE* is explicitly an evaluation of research soundness and *not* its perceived importance. Since its introduction in 2006, *PLoS ONE*'s growth has been astronomical. In 2011, 13,798 articles were published (70% acceptance rate) making it the largest journal in the world. Given the disregard for importance in the review process, one might surmise that *PLoS ONE*'s impact factor would be quite low. In fact, its 2011 impact factor was an impressive 4.41. This put it in the top 25% of general biological science journals, and nearly as high as *Psychological Science* (4.7). This casts further doubt on reviewers' ability to predict importance (Gottfredson, 1978), or at least one indicator of importance - citation impact. With a publishing model focused on soundness, negative results and replications are more publishable, and the journal identity is not defined as publishing research that is otherwise unpublishable.

*Lowering or removing the barrier for publication.* A more radical fix than the *PLoS ONE* model is to discard publishing as a meaningful incentive. How? Make it trivial to publish. The peer review process presently serves as both gatekeeper and evaluator. Post-publication peer review can separate these concepts by letting the author decide when to publish. Then, peer review operates solely as an evaluation mechanism (Armstrong, 1997; Nosek & Bar-Anan, 2012; Smith, 1999). Nosek and Bar-Anan (2012) provide in-depth discussion for how this is achievable by embracing digital journals and public repositories, and by restructuring the review process. Successful models already exist such as arXiv, the public repository for physics and other fields (http://arxiv.org; see also http://ssrn.com/ and http://repec.org/). Authors submit their manuscripts to arXiv making them publicly available to the physics community. Peer review – through the "typical" journals – occurs independently of disseminating manuscripts through the repository. If physicists want to wait for peer review to determine everything they read, they can still do so. But most physicists use arXiv to keep up-to-date on what other laboratories are doing in their specialty.

By making it trivial to publish, the act itself is no longer much of an incentive. Anyone can publish. The incentives would then shift to evaluation of the research and its impact on future research (i.e., its contribution to cumulating knowledge). Also, the priorities in the peer review process would shift from assessing whether the manuscript should be published to whether the ideas should be taken seriously and how they can be improved (Nosek & Bar-Anan, 2012). Further, this would remove a major barrier to publishing replications and negative results if and when they occur. The only barrier left would be the authors' decision of whether it is worthwhile to write up a report at all.

Finally, this change would alter the mindset that publication is the end of the research process. In the present system, is easy to perceive the final step in research occurring when the published article is added to one's vita. That is the incentive of publication, but not of knowledge building. Knowledge building incentives are satisfied when the research has impact on new investigations. By reducing the value of publication, the comparative value of having impact on other research increases (see Nosek & Bar-Anan, 2012 for a detailed discussion and addressing of common concerns about the impact of moving to a post-publication peer review model).

**The Ultimate Solution: Opening Data, Materials, and Workflow**

Implementing the strategies in the previous section will shift the incentives toward more efficient knowledge accumulation. They do not, however, address the core factor that led Motyl and Nosek to conduct a replication in the opening anecdote – accountability. Science is a distributed, non-hierarchical system. As noted by Nosek and Bar-Anan (2012), "Open communication among scientists makes it possible to accumulate a shared body of knowledge. … Individual scientists or groups make claims and provide evidence for those claims. The claims and evidence are shared publicly so that others can evaluate, challenge, adapt, and reuse the methods or ideas for additional investigation. … science makes progress through the open, free exchange of ideas and evidence" (msp. 3). Openness provides scientists confidence in the claims and evidence provided by other scientists. Further, reputation enhancement is a primary mechanism for reward in unstructured contribution systems. Scientists gain and lose status by their public contributions to scientific progress. As such, public reputation management is the primary lever for promoting accountability in academic science.

In present research practice, openness occurs almost entirely through a single mechanism – the journal article. Buckheit and Donoho (1995) suggested that "a scientific publication is not the scholarship itself, it is merely advertising of the scholarship" to emphasize how much of the actual research is opaque to readers. For the objective of knowledge accumulation, the benefits of openness are substantial. Openness increases accountability (Lerner & Tetlock, 1999); makes it easier to share, adapt, extend, and critique methods, materials, analysis scripts, and data; can eliminate the file-drawer effect; and can improve the potential for identifying and correcting errors (Ioannidis & Khoury, 2011; Ioannidis & Panagiotou, 2011; Schooler, 2011; Stodden, 2011).

Three areas of scientific practice – data, methods and tools, and workflow – are largely closed in present scientific practices. Increasing openness in each of them would substantially improve scientific progress.

*Open data*. With the massive growth in data and increased ease of making it available, calls for open data as a standard practice are occurring across all of the sciences (Freese, 2007; King, 2006, 2007; Schofield et al., 2009; Stodden, 2011; Wicherts, 2011; Wicherts & Bakker, 2012). For example, the Human Genome Project acknowledges their principle of rapid, unrestricted release of prepublication data as a major factor for its enormous success in spurring scientific publication and progress (Lander et al., 2001). Arguments for open data cite the ability to confirm, critique, or extend prior research (Smith, Budzieka, Edwards, Johnson, & Bearse, 1986; Wicherts, Borsboom, Kats, & Molenaar, 2006; Wolins, 1962), opportunity to reanalyze prior data with new techniques (Bryant & Wortman, 1978; Hedrick, Boruch, & Ross, 1978; Nosek & Bar-Anan, 2012; Poldrack et al., 2011; Stock & Kulhavy, 1989), increased ability to aggregate data across multiple investigations for improved confidence in research findings (Hrynaszkiewicz, 2010; Rothstein, Sutton, & Borenstein, 2006; Yarkoni et al., 2010), the opportunity for novel methodologies and insights through aggregation and big data (Poldrack et al., 2011), and that openness and transparency increases credibility of science and the findings (Vision, 2010).

The concerns about credibility may be well-founded. In one study, only 27% of psychologists shared at least some of their data upon request for confirming the original results even though APA ethics policies required data sharing for such circumstances (Wicherts et al., 2006; see also Pienta, Gutmann, & Lyle, 2009). Further, Wicherts, Bakker, and Molenaar (2011) found that reluctance to share published data was associated with weaker evidence against the null hypothesis, and more apparent errors in statistical analysis – particularly those that made a difference for statistical significance. This illustrates the conflict between personal interests and scientific progress – the short-term benefit of avoiding identification of one's errors dominated the long-term cost of those errors remaining in the scientific literature.

The rate of errors in published research is unknown, but a study by Bakker and Wicherts (2011) is breathtaking. They reviewed 281 articles and found that 15% contained statistical conclusions that were incorrect – reporting a significant result ($p < .05$) that was not, or vice versa. Their investigation could only catch statistical errors that were detectable in the articles themselves. Errors can also occur in data coding, data cleaning, data analysis, and result reporting. None of those can be detected with only the summary report. For example, in a study looking at sample mix-ups in genome-wide association studies found evidence that every single original dataset examined had at least one sample mix-up error, that the total error rate was 3%, and the worst performing paper – published in a highly prestigious outlet – had 23% of its samples categorized erroneously (Westra et al., 2011). Further, correcting these errors had a substantial impact on improving the sensitivity of identifying markers in the datasets.

Making data openly available increases likelihood of finding and correcting errors, and ultimately improving reported results. Simultaneously, it improves the potential for aggregation of raw data for research synthesis (Cooper, Hedges, & Valentine, 2009), it presents opportunities for applications with the same data that may not have been pursued by the original authors, and it creates a new opportunity for citation credit and reputation-building (Piwowar, 2011; Piwowar, Day, & Fridsma, 2007). Researchers who create useful datasets can be credited for the contribution beyond their own uses of the data.

Movement toward open data is occurring rapidly. Many infrastructure projects are making it easier to share data. There are field-specific options such as OpenfMRI (http://www.openfmri.org/; Poldrack et al., 2011), INDI (http://fcon_1000.projects.nitrc.org/) and OASIS (http://www.oasis-brains.org/) for neuroimaging data. And, there are field-general options such as the Dataverse Network Project (http://thedata.org/), and Dryad (http://datadryad.org/). Some journals are beginning to require data deposit as a condition of publication (Al-Sheikh-Ali, Qureshi, Al-Mallah, & Ioannidis, 2011). Likewise, funding agencies and professional societies are encouraging or requiring data availability post-publication (American Psychological Association, 2010; National Institutes of Health, 2003; National Science Foundation, 2011; PLoS ONE, n.d.).

Of course, while some barriers to sharing are difficult to justify – such as concerns that others might identify errors – others are reasonable (Smith et al., 1986; Stodden, 2010; Wicherts & Bakker, 2012). Researchers may not have a strong ethic of data archiving for past research – it may simply not be available anymore. For available data, many times it is not formatted for easy comprehension and sharing. Preparing it takes additional time (though much less so if the researcher plans to share the data from the outset of the project). Further, there are exceptions for blanket openness such as inability to assure confidentiality of participant identities, legal barriers (e.g., copyright), and occasions in which it is reasonable to delay openness – such as when data collection effort is intense and the dataset is to be the basis for multiple research projects (American Psychological Association, 2010; National Institutes of Health, 2003; National Science Foundation, 2011). The key point is that these are exceptions. Default practice can shift to openness while guidelines are developed for the justification to keep data closed or delay its release (Stodden, 2010).

*Open methods and tools*. Open data allows confirmation, extension, critique and improvement of research already conducted. Opening methods has the same effect and also facilitates progress in reuse, adaptation, and extension for new research (Schofield et al., 2009). In particular, open methodology facilitates replication and paradigm-driven research. Published reports of methodologies often lack sufficient detail to conduct a replication (Donoho et al., 2009; Stodden, 2011). At best, the written report is the authors' understanding of what is critical for the methodology. However, there are many factors that could be important but go unmentioned – for example, the temperature of the room for data collection, the identities of the experimenters, the time of day for data collection, or whether instructions were delivered verbally or in written form. Moreover, in paradigm-driven research, changes to the methodology are ideally done by design, not by accident. The likelihood of replicating and extending a result is stronger if the original materials are re-used and adapted rather than reinvented based on the new researchers' understanding of the original researchers' written description.

Authors cannot identify and report every detail that may be important in a method, but many more parts of the methodology can be shared outside of the report itself. For example, it is easy to create a video of the experimental setting and conduct a simulation of the procedure for posting on the Internet. Figshare (http://figshare.com/) offers a repository for data and methods or materials for private archiving or public sharing. Further, the Open Science Framework (http://openscienceframework.org/) is a web-based project management framework for documenting and archiving research materials, analysis scripts or data, and empowers the user to keep the materials private or make them public.

Presently, only the scientific report is cited and valued. Openness with data, methods, and tools makes them citable contributions (Mooney, 2011; Piwowar, Day, & Fridsma, 2007; http://www.data-pass.org/citations.html). Contributing data or methods that are the basis for multiple investigations provides reputation enhancement for the originator of the resources. Vitas can include citations to the articles, datasets, methods, scripts, and tools that are each independently contributing to knowledge accumulation (Altman & King, 2007). Also, the ready availability of these materials will accelerate productivity by eliminating the need to recreate or reinvent them. Further, reinvention based on another's description of methods is a risk factor for introducing unintended differences between the original and replicated methodology.

*Open workflow*. Given that academic science is a largely public institution funded by public money, it is surprising that there is so little transparency and accountability for the research process. Beyond the published reports, science operates as a "trust me" model that would be seen as laughably quaint for ensuring responsibility and accountability in state or corporate governance.

In some areas of science, however, it is understood that transparency in the scientific workflow underlies credibility and accuracy. For example, http://clinicaltrials.gov/ is an NIH-sponsored study registry for clinical trials. In 2005, the *International Committee of Medical Journal Editors* started requiring authors to register their randomized controlled trials into a registry prior to data collection as a condition for publication. Companies sponsoring trials have an obvious financial conflict of interest for the outcome of the research. A registry makes it more difficult to hide undesired outcomes. Indeed, using registry data, Mathieu, Boutron, Moher, Altman, and Ravaud (2009) found that 31% of adequately registered trials showed discrepancies between the registered and published outcomes. For those in which the nature of the discrepancies could be assessed, 82% of them favored reporting statistically significant results.

Of course, money is not the only source of conflict of interest. Scientists are invested in their research outcomes via their interests, beliefs, ego, and reputation. Some outcomes may be more desirable than others – particularly when personal beliefs or prior claims are at stake. Those desires may translate into design, analysis and reporting decisions that systematically bias the accuracy of what is reported, even without realizing that it is occurring (Kunda, 1990; Mullen, Bauman, & Skitka, 2003). Public documentation of a laboratory's research process makes these practices easier to detect and could reduce the likelihood that they will occur at all (Bourne, 2010). Further, registration of studies prior to their completion solves one aspect of the file-drawer effect – knowing what research was done even if it does not get published (Schooler, 2011).

An obvious concern about transparency of workflow is that researchers are not interested in most of the details of what goes on in other laboratories. Indeed, while advocating this strongly, the present authors do not expect that we would routinely look at the details of other laboratory operations. However, there are occasions for which access would be useful. For example, when we are inspired by another researcher's work and aim to adapt it for our research purposes, we often need more detail than is provided in the summary reports. Access to the materials and workflow will be very useful in those cases. Further, while we do not care to look at the public data about U.S. government expenditures ourselves (http://www.data.gov/), we are pleased with the transparency and the fact that someone can look. Indeed, much as investigative journalism provides accountability for government practice, with open workflow, new contributors to science might emerge who evaluate the knowledge accumulation process rather than produce it, and are valued as such.

Finally, using a registry in an open workflow can clarify whether a finding resulted from a confirmatory test of a strong *a priori* prediction or was a discovery in the course of conducting the research. The current default practice is to tell a good story by reporting findings as if the research had been planned that way (Bem, 2003). However, even if we intend to disclose confirmation versus discovery, our recollection of the project purpose may not be the same as the project purpose when it began. People reconstruct the past through the lens of their present (Schacter, 2001). People are more likely to presume what they know now was how they conceived it at the beginning (Christensen-Szalanski & Willham, 1991; Fischoff, 1977; Fischoff & Beyth, 1975). Without a registry for accountability, findings may be genuinely and confidently espoused as confirmatory tests of prior predictions when they are written for publication. However, discoveries are more likely to leverage chance than are confirmatory tests. What appears to be "what we learned" could be "what chance told us." The point of making a registry available is not to have *a priori* hypotheses for all projects and findings; it is to clarify when there was one and when there was not. When it is a discovery, acknowledge it as a discovery. As Tukey summarized (1977):

> Once upon a time statisticians only explored. Then they learned … to confirm a few things exactly, each under very specific circumstances. As they emphasized exact confirmation, their techniques inevitably became less flexible. The connection of the most used techniques with past insights was weakened. Anything to which a confirmatory procedure was not explicitly attached was decried as "mere descriptive statistics", no matter how much we had learned from it (p. vii).

Discovery is critical for science because learning occurs by having assumptions violated. Strong narratives focusing on what was learned are useful communication devices, and simple disclosures of how it was learned are useful accuracy devices.

## Conclusion

We titled this article "Scientific Utopia" self-consciously. The suggested revisions to scientific practice are presented idealistically. The realities of implementation and execution are messier than their conceptualization. Science is the best available method for cumulating knowledge about nature. Even so, scientific practices can be improved to enhance the efficiency of knowledge building. The present article outlined changes to address a conflict of interest for practicing scientists – the rewards of getting published that are independent of the accuracy of the findings that are published. Some of these changes are systemic and require cultural, institutional, or collective change. But others can emerge "bottom-up" by scientists altering their own practices.

We, the present authors, would like to believe that our motivation to do good science would overwhelm any decisions that prioritize publishability over accuracy. However, publishing is a central, immediate, and concrete objective for our career success. This makes it likely that we will be influenced by self-serving reasoning biases despite our intentions. The most effective remedy available for immediate implementation is to make our scientific practices transparent. Transparency can improve our practices even if no one actually looks, simply because we know that someone *could* look.

Existing technologies allow us to translate some of this ideal into practice. We make our unpublished manuscripts available at personal webpages (e.g., http://briannosek.com/) and public repositories (http://ssrn.com/). We make our study materials and tools available at personal web pages (e.g., http://people.virginia.edu/~msm6sw/materials.html; http://people.virginia.edu/~js6ew/). We make data available through the Dataverse Network (e.g., http://dvn.iq.harvard.edu/dvn/dv/bnosek), and we are contributing to the design and construction of the Open Science Framework for comprehensive management and disclosure of our scientific workflow (http://openscienceframework.org/). Opening our research process will make us feel accountable to do our best to get it right; and, if we do not get it right, to increase the opportunities for others to detect the problems and correct them. Openness is not needed because we are untrustworthy; it is needed because we are human.

**References**

Aldhous, P. (2011). Journal rejects studies contradicting precognition. *New Scientist*. Retrieved from: http://www.newscientist.com/article/dn20447-journal-rejects-studies-contradicting-precognition.html

Alsheikh-Ali, A. A., Qureshi, W., Al-Mallah, M. H., & Ioannidis, J. P. (2011). Public availability of published research data in high-impact journals. *PLoS ONE, 6*, e24357. doi:10.1371/journal.pone.0024357

Altman, M., & King, G. (2007). A proposed standard for the scholarly citation of quantitative data. *D-Lib Magazine, 13(3/4)*. http://gking.harvard.edu/files/abs/cite-abs.shtml

American Psychological Association (2011). *Publication Manual of the American Psychological Association, Sixth Edition American Psychological Association*. Washington, DC: American Psychological Association.

Armstrong, J. S. (1997). Peer review for journals: Evidence on quality control, fairness, and innovation. *Science and Engineering Ethics, 3*, 63-84.

Bacon, R. (1267/1859). Fr. Rogeri Bacon Opera quædam hactenus inedita. Vol. I. containing I.--Opus tertium. II.--Opus minus. III.--Compendium philosophiæ. Longman, Green, Longman and Roberts. Retrieved from: http://books.google.com/books?id=wMUKAAAAYAAJ

Bakan, D. (1966). The test of significance in psychological research. *Psychological Bulletin, 66*, 423-437.

Bakker, M., & Wicherts, J. M. (2011). The (mis)reporting of statistical results in psychology journals. *Behavior Research*. doi: 10.3758/s13428-011-0089-5

Ball, P. (2005). Index aims for fair ranking of scientists. *Nature, 436,* 900.

Bastardi, A., Uhlmann, E. L., & Ross, L. (2011). Wishful thinking: Beliefs, desire, and the motivated evaluation of scientific evidence. *Psychological Science, 22*, 731-732. doi: 10.1177/0956797611406447

Begley, C. G., & Ellis, L. M. (2012). Raise standards for preclinical cancer research. *Nature, 483*, 531-533.

Bem, D. J. (2003). Writing the empirical journal article. In J. M. Darley, M. P. Zanna, & H. L. Roediger III (Eds.), *The compleat academic: A career guide (pp. 171–201)*. Washington, DC: American Psychological Association.

Bersoff, D. M. (1999). Why good people sometimes do bad things: Motivated reasoning and unethical behavior. *Personality and Social Psychology Bulletin, 25*, 28-39.

Boiney, L. G., Kennedy, J., & Nye, P. (1997). Instrumental bias in motivated reasoning: More when more is needed. *Organizational Behavior and Human Decision Processes, 72*, 1-24.

Bornmann, L., Mutz, R., & Daniel, H.-D. (2010). A reliability-generalization study of journal peer reviews: a multilevel meta-analysis of inter-rater reliability and its determinants. *PloS ONE, 5(12)*, e14331. doi:10.1371/journal.pone.0014331

Bouwmeester, W., Zuithoff, N. P. A., Mallett, S., Geerlings, M. I., Vergouwe, Y., et al. (2012). Reporting and methods in clinical prediction research: A systematic review. *PLoS Medicine, 9*, e1001221. doi:10.1371/journal.pmed.1001221

Braude, S. E. (1979). *ESP and psychokinesis*. A philosophical examination. Philadelphia: Temple University Press.

Bryant, F. B., & Wortman, P. M. (1978). Secondary analysis: The case for data archives. *American Psychologist, 33,* 381-387. doi: 10.1037/0003-066X.33.4.381

Buckheit, J. B., & Donoho, D. L. (1995). *WaveLab and Reproducible Research*. Department of Statistics, Stanford University, Technical Report 474. Retrieved from: http://www-stat.stanford.edu/~wavelab/Wavelab_850/wavelab.pdf

Budd, J. M., Sievert, M., & Schultz, T. R. (1998). Phenomena of retraction: reasons for retraction and citations to the publications. *The Journal of the American Medical Association, 280(3)*, 296-297. Retrieved from http://www.ncbi.nlm.nih.gov/pubmed/9676689

Butler, K. M., McDaniel, M. A., Dornburg, C. C., Price, A. L., & Roediger, H. L., III (2004). Age differences in veridical and false recall are not inevitable: The role of frontal lobe function. *Psychonomic Bulletin and Review, 11*, 921-925.

Carpenter, S. (2012). Psychology's Bold Initiative. *Science, 335*, 1558-1560.

Christensen-Szalanski, J. J. J., & Willham, C. F. (1991). The hindsight bias: A meta-analysis. *Organizational Behavior and Human Decision Processes, 48*, 147-168.

Cicchetti, D. V. (1991). The reliability of peer review for manuscript and grant submissions: A cross-disciplinary investigation. *Behavioral and Brain Sciences, 14*, 119-135.

Cohen , J. (1962). The statistical power of abnormal-social psychological research: A review. *Journal of Abnormal and Social Psychology, 65*, 145-153.

Cohen , J. (1969). *Statistical power analysis for the behavioral sciences*. San Diego, CA: Academic Press.

Cohen, J. (1992). A power primer. *Psychological Bulletin, 112*, 155-159.

Cohen, J. (1994). The earth is round (p < .05). *American Psychologist, 49*, 997-1003. doi: 10.1037/0003-066X.49.12.997

Collins, H. M. (1985). *Changing order*. London: Sage.

Ditto, P. H., & Lopez, D. F. (1992). Motivated skepticism: Use of differential decision criteria for preferred and nonpreferred conclusions. *Journal of Personality and Social Psychology, 63*, 568-584. doi:10.1037/0022-3514.63.4.568

Donoho, D. L., Maleki, A., Rahman, I. U., Shahram, M., & Stodden, V. (2009). Reproducibility research in computational harmonic analysis. *Computing, Science, & Engineering, 11*, 8-18.

Edge, H. (1985). The problem is not replication. In B. Shapin & L. Coly (Eds.), *The repeatability problem in parapsychology (pp. 53–64)*. New York: The Parapsychology Foundation.

Elms, A. C. (1975). The crisis of confidence in social psychology. *American Psychologist, 30*, 967–976.

Fanelli, D. (2009). How many scientists fabricate and falsify research? A systematic review and meta-analysis of survey data. *PLoS ONE, 4(5)*, 1-11.

Fanelli, D. (2010a). "Positive" results increase down the hierarchy of the sciences. *PLoS ONE, 5(4)*, e10068. doi:10.1371/journal.pone.0010068

Fanelli, D. (2010b). Do pressures to publish increase scientists' bias? An Empirical Support from US States Data. *PLoS ONE, 5(4)*, e10271. doi:10.1371/journal.pone.0010271

Fanelli, D. (2012). Negative results are disappearing from most disciplines and countries. *Scientometrics, 90*, 891-904.

Feyerabend, P. (1975). *Against method*. London, UK: New Left Books.

Fischoff, B. (1977). Perceived informativeness of facts. *Journal of Experimental Psychology: Human Perception and Performance, 3*, 349-358. doi: 10.1037/0096-1523.3.2.349

Fischoff, B., & Beyth, R. (1975). "I knew it would happen" Remembered probabilities of once-future things. *Organizational Behaviour and Human Performance, 13*, 1-16

Freese, J. (2007). Overcoming objections to open-source social science. *Sociological Methods & Research, 36*, 220-226.

Frick, R. W. (1996). The appropriate use of null hypothesis testing. *Psychological Methods, 1*, 379-390.

Gallo, D. A. (2010). False memories and fantastic beliefs: 15 years of the DRM illusion. *Memory and Cognition, 38*, 833-848. doi:10.3758/MC.38.7.833

Gardner, W., Lidz, C. W., & Hartwig, K. C. (2005). Authors' reports about research integrity problems in clinical trials. *Contemporary Clinical Trials, 26(2)*, 244-251.

Gawande, A. (2009). *The Checklist Manifesto*. New York, NY: Metropolitan Books.

Giner-Sorolla, R. (2012). Science or art? How esthetic standards grease the way through the publication bottleneck but undermine science. Unpublished manuscript.

Gottfredson, S. D. (1978). Evaluating psychological research reports: Dimensions, reliability, and correlates of quality judgments. *American Psychologist, 33*, 920-934.

Greenwald, A. G. (1975). Consequences of prejudice against the null hypothesis. *Psychological Bulletin, 82*, 1–20.

Hackett, B. (2005). Essential tensions: Identity, control, and risk in research. *Social Studies of Science, 35(5)*, 787–826. doi:10.1177/0306312705056045.

Haslam, N., Ban, L., Kaufmann, L., Loughnan, S., Peters, K., Whelan, J., & Wilson, S. (2008). What makes an article influential? Predicting impact in social and personality psychology. *Scientometrics, 76*, 169-185.

Hedrick , T. E. Boruch , R. F. Ross , J. (1978). On ensuring the availability of evaluation data for secondary analysis. *Policy Sciences, 9,* 259-280.

Hewitt, J. K. (2012). Editorial policy on candidate gene association and candidate gene-by-environment interaction studies of complex traits. *Behavior Genetics, 42*, 1-2. doi: 10.1007/s10519-011-9504-z

Hrynaszkiewicz, I. (2010). A call for BMC Research Notes contributions promoting best practice in data standardization, sharing, and publication. *BioMed Central Research Notes, 3,* 235. doi: 10.1186/1756-0500-3-235

Ioannidis, J. P. A. (2005). Why most published research findings are false. *PLoS Medicine, 2*, e124.

Ioannidis, J.P.A., & Khoury, M.J. (2011). Improving validation practices in "omics" research. *Science, 334*, 1230-1232.

Ioannidis, J., & Panagiotou (2011). Comparison of effect sizes associated with biomarkers reported in highly cited individual articles and in subsequent meta-analyses. *Journal of the American Medical Association, 305, 2200*-2210. doi:10.1001/jama.2011.713

Ioannidis, J. P. A., & Trikalinos, T. A. (2007). An exploratory test for an excess of significant findings. *Clinical Trials, 4*, 245–253.

Jasny, B. R., Chin, G., Chong, L., & Vignieri, S. (2011). Again, and again, and again… *Science, 334,* 1225. doi: 10.1126/science.334.6060.1225

John, L., Loewenstein, G., & Prelec, D. (2012). Measuring the prevalence of questionable research practices with incentives for truth-telling. *Psychological Science*, *23*, 524-532. doi: 10.1177/0956797611430953

Kaiser, C. R. (2012). Campaign for real data. *Dialogue, 26*, 8–10.

Kerr, N. L. (1998). HARKing: Hypothesizing after the results are known. *Personality and Social Psychology Review, 2*, 196–217.

King, G. (2006). Publication, publication. *PS-Political Science & Politics, 39*, 119-125.

King, G. (2007). An introduction to the dataverse network as an infrastructure for data sharing. *Sociological Methods & Research, 36*, 173-199.

Kuhn, T.S. (1962). *The structure of scientific revolutions*. Chicago, IL: University of Chicago Press.

Kunda, Z. (1990). The case for motivated reasoning. *Psychological Bulletin, 108,* 480-498. doi:10.1037/0033-2909.108.3.480

Lakatos, I. (1978). *The methodology of scientific research programmes: Philosophical papers Volume 1*. Cambridge: Cambridge University Press.

Lander, E. S, Linton, L. M., Birren, B., Nusbaum, C., Zody, M. C., et al. (2001). Initial sequencing and analysis of the human genome. *Nature, 409,* 860-921.

Lerner, J., & Tetlock, P. E. (1999). Lerner, J. & Tetlock, P.E. (1999). Accounting for the effects of accountability. Psychological Bulletin, 125, 255-275. doi: 10.1037/0033-2909.125.2.255

Liberman, N., & Trope, Y. (1998). The role of feasibility and desirability considerations in near and distant future decisions: A test of temporal construal theory. *Journal of Personality and Social Psychology, 75*, 5-18.

Lindsay, R. M., & Ehrenberg, A. S. C. (1993). The design of replicated studies. *The American Statistician, 47*, 217–228.

Lord, C. G., Ross, L., & Lepper, M. R. (1979). Biased assimilation and attitude polarization: The effects of prior theories on subsequently considered evidence. *Journal of Personality and Social Psychology, 37*, 2098-2109. doi:10.1037/0022-3514.37.11.2098

Lykken, D. T. (1968). Statistical significance in psychological research. *Psychological Bulletin, 70*, 151–159.

Madden, C. S., Easley, R. W., & Dunn, M. G. (1995). How journal editors view replication research. *Journal of Advertising, 24*, 78–87.

Mahoney, M. J. (1985). Open exchange and epistemic process. *American Psychologist, 40*, 29–39.

Markman, A. B., & Brendl, C. M. (2005). Constraining theories of embodied cognition. *Psychological Science, 16*, 6-10. DOI: 10.1111/j.0956-7976.2005.00772.x

Marsh, H. W., Ball, S.  (1989). The peer review process used to evaluate manuscripts submitted to academic journals : Interjudgmental reliability. *The Journal of Experimental Education, 57(2)*, 151-169.

Marsh, H. W., Jayasinghe, U. W., & Bond, N. W. (2008). Improving the peer review process for grant applications: Reliability, validity, bias, and generalizability. *American Psychologist, 63*, 160-168. Doi: 10.1037/0003-066X.63.3.160

Martin, B. (1992). Scientific fraud and the power structure of science. *Prometheus, 10(1)*, 83–98. DOI:10.1080/08109029208629515

Martinson, B. C., Anderson, M. S., & Devries, R. (2005). Scientists behaving badly. *Nature, 435*, 737-738.

Mathieu, S., Boutron, I., Moher, D., Altman, D. G., & Ravaud, P. (2009). Comparison of registered and published primary outcomes in randomized controlled trials. *Journal of the American Medical Association, 302,* 977-984. doi:10.1001/jama.2009.1242

McGuire, W. J. (2004). A perspectivist approach to theory construction. *Personality and Social Psychology Review, 8*, 173-182.

Merton, R. K. (1942). Science and technology in a democratic order. *Journal of Legal and Political Sociology, 1*, 115–126.

Merton, R. K. (1973). *The Sociology of Science, Theoretical and Empirical Investigations*. The University of Chicago Press, Chicago.

Miller, A. C., & Serzan, S. L. (1984). Criteria for identifying a refereed journal. *The Journal of Higher Education, 55*, 673–699.

Mooney, H. (2011). Citing data sources in the social sciences: do authors do it? *Learned Publishing, 24*, 99-108.

Motyl, M., & Nosek, B. A. (2012). Unpublished data.  University of Virginia.

Mullen, E., Bauman, C. W., & Skitka, L. J. (2003). Avoiding the pitfalls of politicized psychology. *Analyses of Social Issues and Public Policy, 3*, 171-176.

National Institutes of Health (2003). Final NIH Statement on Sharing Research Data. In *NIH Data Sharing Policy*. Retrieved May 17, 2012, from http://grants.nih.gov/grants/guide/notice-files/NOT-OD-03-032.html.

National Science Foundation (2011). Dissemination and Sharing of Research Results. In *Grant Proposal Guide*. Retrieved May 17, 2012, from http://www.nsf.gov/pubs/policydocs/pappguide/nsf11001/aag_6.jsp.

Neuliep, J. W., & Crandall, R. (1990). Editorial bias against replication research. *Journal of Social Behavior and Personality, 5*, 85–90.

Neuliep, J. W., & Crandall, R. (1993). Reviewer bias against replication research. *Journal of Social Behavior and Personality, 8*, 21–29.

Nickerson, R. S. (1998). Confirmation bias: A ubiquitous phenomenon in many guises. *Review of General Psychology, 2*, 175-220.

Nickerson, R. S. (2000). Null hypothesis significance testing: A review of an old and continuing controversy. *Psychological Methods, 5*, 241-301. DOI: 10.1037/1082-989X.5.2.241

Nosek, B. A. & Bar-Anan, Y. (2012). Scientific Utopia: I. Opening scientific communication. *Psychological Inquiry*.

Nosek, B. A., Graham, J., Lindner, N. M., Kesebir, S., Hawkins, C. B., Hahn, C., Schmidt, K., Motyl, M., Joy-Gaba, J., Frazier, R., & Tenney, E. R. (2010). Cumulative and career-stage citation impact of social-personality programs and their members. *Personality and Social Psychology Bulletin, 36*, 1283-1300.

Nosek, B. A., Smyth, F. L., Hansen, J. J., Devos, T., Lindner, N. M., Ranganath, K. A., Smith, C. T., Olson, K. R., Chugh, D., Greenwald, A. G., & Banaji, M. R. (2007). Pervasiveness and correlates of implicit attitudes and stereotypes. *European Review of Social Psychology, 18*, 36-88.

Oakes, M. (1986). *Statistical inference: A commentary for the social and behavioral sciences*. New York: Wiley.

Open Science Collaboration. (2012a). *Possible interpretations of a failure to replicate*. Retrieved from: https://docs.google.com/document/d/10x-uzlQ2vIQgsHNum2U9VC0M289lXZozR41MeHqFy2M/

Open Science Collaboration. (2012b). Replication value. Unpublished manuscript.

Osherovich, L. (2011). Hedging against academic risk. *Science-Business eXchange, 4(15)*. doi:10.1038/scibx.2011.416

Ostriker, J. P., Holland, P. W., Kuh, C. V., & Voytuk, J. A. (2009). *A guide to the methodology of the National Research Council assessment of doctorate programs*. Washington, DC: National Academic Press.

Peters, D. P., & Ceci, S. J. (1982). Peer-review practices of psychological journals: The fate of published articles, submitted again. *Behavioral and Brain Sciences, 5*, 187–255.

Petty, R. E., Fleming, M. A., & Fabrigar, L. R. (1999). The Review Process at PSPB: Correlates of Interreviewer Agreement and Manuscript Acceptance. *Personality and Social Psychology Bulletin, 25*, 188-203. doi:10.1177/0146167299025002005

Phillips, N. (2007). Citation counts, prestige measurement, and graduate training in social psychology. *Dialogue*, *22*(2), 24-26.

Pienta, A. M., Gutmann, M. P., & Lyle, J. (2009). *Research Data in the Social Sciences: How Much is Being Shared?* Paper presented at the Research Conference on Research Integrity, Niagara Falls, NY.

Piwowar, H. A. (2011). Who shares? Who doesn't? Factors associated with openly archiving raw research data. *PLoS ONE, 6*, e18657.

Piwowar, H. A., Day, R. S., & Fridsma, D. B. (2007). Sharing detailed research data is associated with increased citation rate. *PLoS ONE, 2*, e308.

PLoS ONE (n.d.). Sharing of Materials, Methods, and Data. In *PLoS Editorial and Publishing Policies*. Retrieved May 17, 2012, from http://www.plosone.org/static/policies.action#sharing.

Popper, K. (1934/1992). *The Logic of Scientific Discovery*. New York: Routledge.

Prinz, F., Schlange, T. & Asadullah, K. (2011). Believe it or not: how much can we rely on published data on potential drug targets? *Nature Reviews Drug Discovery, 10*, 712-713.

Proffitt, D. R. (2006). Embodied perception and the economy of action. *Perspectives on Psychological Science, 1*, 110-122. DOI: 10.1111/j.1745-6916.2006.00008.x

Pyszczynski, T., & Greenberg, J. (1987). Perspectives on social inference: A biased hypothesis-testing model. In L. Berkowitz (Ed.), *Advances in Experimental Social Psychology Vol. 20*, (297-340). doi: 10.1016/S0065-2601(08)60417-7

Redman, B. K., Yarandi, H. N., & Merz, J. F. (2008). Empirical developments in retraction. *Journal of Medical Ethics, 34(11)*, 807-809. doi:10.1136/jme.2007.023069

Reid, L. N., Soley, L. C., & Wimmer, R. D. (1981). Replication in advertising research: 1977, 1978, 1979. *Journal of Advertising, 10*, 3-13.

Reyna, V.F., & Brainerd, C.J. (1995). Fuzzy trace theory: An interim synthesis. *Learning and Individual Differences, 7*, 1–75.

Roediger, H. L. (2012). Psychology's woes and a partial cure: The value of replication. *APS Observer, 25(2)*.

Roediger, H. L., & McDermott, K. B. (1995). Creating false memories: Remembering words not presented in lists. *Journal of Experimental Psychology: Learning, Memory, and Cognition, 21*, 803-814.

Rosenthal, R. (1979). The file drawer problem and tolerance for null results. *Psychological Bulletin, 86,* 638-641. doi: 10.1037/0033-2909.86.3.638

Rosenthal, R. (1991). Replication in behavioral research. In J. W. Neuliep (Ed.), *Replication research in the social sciences* (pp. 1–39). Newbury Park: Sage.

Rosnow, R. L., & Rosenthal, R. (2009). Effect sizes: Why, when, and how to use them. *Journal of Psychology, 217*, 6-14. DOI: 10.1027/0044-3409.217.1.6

Rothstein, H. R., Sutton, A. J., & Borenstein, M. (Eds.). (2006). *Publication bias in meta-analysis*. New York: John Wiley & Sons, Ltd.

Rozeboom, W. W. (1960). The fallacy of the null-hypothesis significance test. *Psychological Bulletin, 57*, 416-428. DOI: 10.1037/h0042040

Sabra, A. I., trans. (1021/1989), *The Optics of Ibn al-Haytham. Books I–II–III: On Direct Vision*. English Translation and Commentary. 2 vols, Studies of the Warburg Institute, vol. 40, London: The Warburg Institute, University of London, ISBN 0-85481-072-2.

Schacter, D. L. (2001). *The seven sins of memory. New York: Houghton Mifflin.*

Schacter, D. L., Israel, L., & Racine, C. (1999). Suppressing false recognition in younger and older adults: The distinctiveness heuristic. *Journal of Memory & Language, 40*, 1-24.

Schmidt, S. (2009). Shall we really do it again? The powerful concept of replication is neglected in the social sciences. *Review of General Psychology, 13*, 90-100.

Schofield, P. N., Bubela, T., Weaver, T., Portilla, L., Brown, S. D., Hancock, J. M., Einhorn, D., Tocchini-Valentini, G., Hrabe de Angelis, M., & Rosenthal, N. (2009). Post-publication sharing of data and tools. *Nature, 461*, 171–173.

Schooler, J. W. (2011). Unpublished results hide the decline effect. *Nature, 470*, 437.

Schroter, S., Black, N., Evans, S., Godlee, F., Osorio, L., & Smith, R. (2008). What errors do peer reviewers detect, and does training improve their ability to detect them? *Journal of the Royal Society of Medicine, 101*, 507-514.

Scientific method (n.d.). Retrieved February 13, 2012 from Wikipedia: http://en.wikipedia.org/wiki/Scientific_method

Simmons, J. P., Nelson, L. D., & Simonsohn, U. (2011). False-positive psychology: Undisclosed flexibility in data collection and analysis allows presenting anything as significant. *Psychological Science, 22*, 1359-1366.

Smith, R. (1999). Opening up BMJ peer review. *BMJ, 318*, 4. DOI: 10.1136/bmj.318.7175.4

Smith, P. C., Budzieka, K. A., Edwards, N. A., Johnson, S. M., & Bearse, L. N. (1986). Guidelines for clean data: Detection of common mistakes. *Journal of Applied Psychology, 71,* 457-460. doi: 10.1037/0021-9010.71.3.457

Sovacool, B. K. (2008). Exploring scientific misconduct: Isolated individuals, impure institutions, or an inevitable idiom of modern science? *Journal of Bioethical Inquiry, 5*, 271-282. doi: 10.1007/s11673-008-9113-6

Sterling, T. D. (1959). Publication decisions and their possible effects on inferences drawn from tests of significance—or vice versa. *Journal of the American Statistical Association, 54*, 30-34.

Sterling, T. D., Rosenbaum, W. L., & Weinkam, J. J. (1995). Publication decisions revisited: The effect of the outcome of statistical tests on the decision to publish and vice versa. *The American Statistician, 49*, 108-112.

Stock, W. A., & Kulhavy, R. W. (1989). Reporting primary data in scientific articles: Technical solutions to a perennial problem. *American Psychologist, 44,* 741-742. doi: 10.1037/0003-066X.44.4.741

Stodden, V. (2010). The Scientific Method in Practice: Reproducibility in the Computational Sciences. MIT Sloan Research Paper No. 4773-10. doi: 10.2139/ssrn.1550193

Stodden, V. (July 2011). Trust your science? Open your data and code. *Amstat News*, 21-22.

Storbeck, J., & Clore, G. L. (2005). With sadness comes accuracy; With happiness, false memory: Mood and the false memory effect. *Psychological Science, 16*, 785-791. doi: 10.1111/j.1467-9280.2005.01615.x

Trope, Y., & Bassok, M. (1982). Confirmatory and diagnosing strategies in social information gathering. *Journal of Personality and Social Psychology, 43*, 22-34. doi:10.1037/0022-3514.43.1.22

Trope, Y., & Liberman, N. (2003). Temporal construal. *Psychological Review, 110*, 403-421.

Tukey, J. W. (1977). *Exploratory Data Analysis*. Reading, MA: Addison-Wesley.

Vision, T. J. (2010). Open data and the social contract of scientific publishing. *BioScience, 60*, 330-331.

Wagenmakers, E. J., Wetzels, R., Borsboom, D., & van der Maas, H. (2011). Why psychologists must change the way they analyze their data: The case of psi. *Journal of Personality and Social Psychology, 100*, 426–432.

Westra, H-. J., Jansen, R. C., Fehrmann, R. S. N., te Meerman, G. J., van Heel, D., Wijmenga, C., & Franke, L. (2011). MixupMapper: correcting sample mix-ups in genome-wide datasets increases power to detect small genetic effects. *Bioinformatics, 27*, 2104-2111.

Whitehurst, G. J. (1984). Interrater agreement for journal manuscript reviews. *American Psychologist, 39*, 22-28. doi:10.1037/0003-066X.39.1.22

Wicherts, J. M. (2011). Psychology must learn a lesson from fraud case. *Nature, 480*, 7. doi:10.1038/480007a

Wicherts, J. M., Bakker, M., & Molenaar, D. (2011). Willingness to share research data is related to the strength of the evidence and the quality of reporting of statistical results. *PLoS One, 6*, e26828. doi:10.1371/journal.pone.0026828

Wicherts, J. M., Borsboom, D., Kats, J., and Molenaar, D. (2006). The poor availability of psychological research data for reanalysis. *American Psychologist, 61*, 726–728. doi:10.1037/0003-066X.61.7.726

Wilkinson, L., & Task Force on Statistical Inference (1999). Statistical methods in psychology journals: Guidelines and explanations. *American Psychologist, 54*, 594–604.

Wolins, L. (1962). Responsibility for raw data. *American Psychologist, 17,* 657-658. doi: 10.1037/h0038819

Wong, P. T. (1981). Implicit editorial policies and the integrity of psychology as an empirical science. *American Psychologist, 36*, 690–691.

Yarkoni, T., Poldrack, R. A., Van Essen, D. C., & Wager, T. D. (2010). Cognitive neuroscience 2.0: Building a cumulative science of human brain function. *Trends in Cognitive Sciences, 14*, 489-496.

Yong, E. (2012). Bad Copy. *Nature, 485*, 298-300.

Young, N. S., Ioannidis, J. P. A., & Al-Ubaydli, O. (2008). Why current publication practices may distort science. *PLoS Medicine, 5*, 1418–1422.

Zuckerman, H., & Merton, R. K. (1971). Patterns of evaluation in science: Institutionalization, structure and functions of the referee system. *Minerva, 9*, 66-100.

**Table. Short list from an assistant professor job search at University of Virginia**

| Current Status | Publications | First Author |
|---|---|---|
| Graduate Student | 8 | 6 |
| Graduate Student | 10 | 4 |
| Graduate Student | 5 | 1 |
| Graduate Student | 4 | 2 |
| Postdoc | 35 | 20 |
| Postdoc | 7 | 3 |
| Postdoc | 8 | 2 |
| Postdoc | 6 | 2 |
| Asst Prof (4 years post PhD) | 8 | 3 |
| Asst Prof (4 years post PhD) | 21 | 12 |
| Asst Prof (4 years post PhD) | 16 | 13 |
| | | |
| Min | 4 | 1 |
| Max | 35 | 20 |
| Mean (Grad Students) | 7 | 3 |
| Mean (Post Docs) | 14 | 7 |
| Mean (Asst Profs) | 15 | 9 |

Note: The job search occurred in the 2000's. The original pool contained more than 100 applications.